\documentclass[preprint,preprintnumbers,amsmath,amssymb,prl]{revtex4}
\usepackage[pdftex]{graphicx}
\usepackage{dcolumn}
\usepackage{bm}
\usepackage{float}
\usepackage{color}
\usepackage{hyperref}
\usepackage{natbib}
\usepackage[T1]{fontenc}
\usepackage{wasysym}
\begin{document}
	\title{\color{blue}Influence of particle size on the thermoresponsive and rheological properties of aqueous poly($N$-isopropylacrylamide) colloidal suspensions}
	\author{Chandeshwar Misra}
	\email{chandeshwar@rri.res.in}
	\affiliation{Soft Condensed Matter Group, Raman Research Institute, C. V. Raman Avenue, Sadashivanagar, Bangalore 560 080, INDIA}
	\author{Sanjay Kumar Behera}
	\email{sanjay@rri.res.in}
	\affiliation{Soft Condensed Matter Group, Raman Research Institute, C. V. Raman Avenue, Sadashivanagar, Bangalore 560 080, INDIA}
	\author{Ranjini Bandyopadhyay}
	\email{ranjini@rri.res.in}
	\affiliation{Soft Condensed Matter Group, Raman Research Institute, C. V. Raman Avenue, Sadashivanagar, Bangalore 560 080, INDIA}
	
	\vspace{0.5cm}
	\date{\today}
	
	\begin{abstract}
		Thermoresponsive poly(N-isopropylacrylamide) (PNIPAM) particles of different sizes are synthesized by
		varying the concentration of sodium dodecyl sulphate (SDS) in a one-pot method. The sizes, size polydispersities and the thermoresponsivity of the PNIPAM particles are characterized by using dynamic light scattering and scanning electron microscopy. It is observed that the sizes of these particles decrease with increase in SDS concentration. Swelling ratios of PNIPAM particles measured from the thermoresponsive curves are observed to increase with decrease in particle size. This observation is understood by minimizing the Helmholtz free energy of the system with respect to the swelling ratio of
		the particles. Finally, the dynamics of these particles in jammed aqueous suspensions are investigated by performing rheological measurements.

		
	\end{abstract}
	
	\maketitle     
	%
	%
	\section{1. Introduction}
	Thermoresponsive poly($N$-isopropylacrylamide) (PNIPAM) particles, suspended in an aqueous medium, can swell and deswell in response to changes in the temperature of the medium \cite{M.Heskins_1968}. PNIPAM particles in aqueous suspensions undergo a reversible volume phase transition above the lower critical solution temperature (LCST) of $\approx$ $34^{\circ}$C \cite{M.Heskins_1968,Y. Hirokawa_1984}. Below the LCST, PNIPAM particles absorb water and swell. Above the LCST, however, they become hydrophobic and expel water which results in dramatic shrinkage in their sizes \cite{H. G. Shild_1992}. As a result  of this thermoresponsive behaviour, PNIPAM particles offer an advantage over many temperature insensitive hard particles like polystyrene and PMMA. The volume fraction of the PNIPAM particles in an aqueous suspension can be tuned by changing the  temperature of the medium  without changing the particle number density \cite{J. Wu_2003, J. Appel_2016}. An important consequence of the deformability and compressibility of soft hydrogel particles is that model hydrogels, such as aqueous PNIPAM suspensions, can be packed well above the random close packing fraction of monodisperse hard spheres, $\phi_{rcp} \approx 0.64$. Due to their thermoresponsive properties and deformability, PNIPAM particles have been widely used in  many applications like drug delivery \cite{L. Dong_1991, C. M. Nolan_2005} and as bio-sensors \cite{M. R. Islam_2014}. Moreover, dense aqueous suspensions of PNIPAM particles can be used as model systems to study the structures and dynamics of the kinetically arrested state of dense colloidal suspensions \cite{P. J. yunker, S. K. Behera_2017}. 
	
	The temperature-dependant phase behaviour of dense aqueous PNIPAM suspensions has been studied extensively \cite{H. senff_1999, V. Carrier_2009, G. Romeo_2010}. It has been reported that below the LCST, PNIPAM particles swell and interact repulsively. The interaction between the particles changes from repulsive to attractive with increase in temperature above the LCST \cite{G. Romeo_2010}. An interesting consequence of this property is that as the temperature is increased well beyond the LCST, the particles aggregate and assemble into a gel phase \cite{Y. Hirose_1987, K. Kratz_2000}. The PNIPAM particles in dense aqueous suspensions can therefore undergo a reversible phase transition from a glassy to a gel-like state while passing through an intermediate liquid-like state.
	
	The thermoresponsive and rheological behaviours of PNIPAM colloidal particles have been widely investigated by varying the stiffness of the PNIPAM particles \cite{K. Kratz_2000, B. S. Martin_2012, P. Voudouris_2013}. This has been achieved experimentally by varying the concentration of the crosslinker $N$,$N^{'}$-methylenebisacrylamide (MBA) during the synthesis process \cite{X. Wu_1994, H. Sinff_200}. It has been reported that stiff PNIPAM particles in dense aqueous suspensions exhibit a phase transition from a liquid to a glassy state while passing through an intermediate crystalline phase with increase in the particle volume fraction. Particles with intermediate stiffnesses, in contrast, transform to a glassy state from a liquid state directly without passing through a crystalline phase \cite{B. S. Martin_2012}. Soft PNIPAM particles suspended in water, however, remain in a liquid phase for the entire range of volume fractions explored in these experiments. The stiffness of the PNIPAM particles significantly affects the mechanical behaviour of these particles in dense aqueous suspensions. It was observed that the plateau of the elastic modulus of dense aqueous PNIPAM suspensions becomes weaker when the stiffness of the particles is decreased \cite{H. Sinff_200}. This is because a higher crosslinking density leads to a higher interparticle force. Another way of changing the particle stiffness is by varying the surfactant sodium dodecyl sulfate (SDS) concentration during the synthesis process \cite{X. Wu_1994, W. Mcphee_1993}. It was reported that an increase in SDS concentration leads to decrease in stiffness and size of the PNIPAM particles in aqueous suspensions \cite{W. Mcphee_1993}. SDS adsorbs on the surface of the primary particles during the nucleation process. This increases the colloidal stability of the particles and thus decreases the size of the particles.

	In this study, we have reported the influence of particle size on the thermoresponsive and rheological properties of PNIPAM particles in aqueous suspensions. The rheology of jammed suspensions of these particles is interesting because of our ability to control the particle stiffnesses, which in turn, alter the packing properties of their dense suspensions. No studies in the literature, to the best of our knowledge, connect particle stiffness, varied by varying SDS concentration, to bulk/macroscopic properties such as rheology.

	In this research study, PNIPAM particles of different sizes were synthesized by varying the sodium dodecyl sulfate (SDS) concentration in the one-pot emulsion polymerization method \cite{X. Wu_1994, S. K. Behera_2017}. The sizes and size distributions of PNIPAM particles were estimated using dynamic light scattering (DLS) and scanning electron microscopy (SEM). The size distribution of PNIPAM particles in aqueous suspensions has been characterized by a dimensionless polydispersity index (PDI). Differential scanning calorimetry (DSC) experiments were performed to characterize the LCSTs of PNIPAM particles of different sizes and stiffnesses in aqueous suspensions. Systematic DLS, SEM and rheological measurements were performed to study the thermoresponsive and rheological behaviors of PNIPAM particles in aqueous suspensions. Crosslinker densities (CLDs) of PNIPAM particles in aqueous suspensions were estimated by using a statistical model for the Helmholtz free energies of these hydrogels. Finally, we performed temperature sweep oscillatory rheological tests to study the thermoresponsive behavior of these suspensions. We have connected the bulk rheology of these suspensions to the stiffnesses of individual particles in aqueous suspensions.

	\section{2. MATERIALS AND METHODS}
	\subsection{2.1 Synthesis of PNIPAM particles by the emulsion polymerization method}
	PNIPAM particles were synthesized by following the one-pot emulsion polymerization method. Particles of different sizes were synthesized by varying the concentration of the surfactant SDS. All the chemicals were purchased from Sigma-Aldrich and used as received without further purification. In the polymerization reaction, $2.8$ g $N$-isopropylacrylamide (NIPAM) ($\geq 99\%$), $0.28$ g $N$,$N^{'}$-methylenebisacrylamide (MBA) ($\geq 99.5\%$) and SDS were dissolved in $190$ mL of Milli-Q water (Millipore Corp.) in a three-necked round-bottomed (RB) flask  attached with a reflux condenser, a magnetic stirrer with heating (Heidolph), a platinum sensor and a nitrogen gas (N$_{2}$) inlet/outlet. The concentration of SDS was varied between 0.01 g/L and 0.6 g/L. The solution was stirred at 600 RPM and purged with N$_{2}$ gas for $30$ min to remove oxygen dissolved in water. The emulsion polymerization reaction was initiated by the addition of 0.112 g of potassium persulphate (KPS) ($ 99.9\%$) dissolved in 10 ml of Milli-Q water after heating the mixture to $70$$^{\circ}$C. The reaction was allowed to proceed for 4 h with a constant stirring speed of 600 RPM. After 4 h, the dispersion was cooled down to room temperature. The dispersion was then purified by 4 successive centrifugations and re-dispersions at speeds of 20,000 to 60,000 RPM for 60 min. Higher centrifugation speeds were used for particles synthesized at higher SDS concentrations and lower speeds were used for particles synthesized at lower SDS concentrations. After the centrigugation, the supernatant was removed and the remaining sample was dried by evaporating the water. A fine powder was prepared by grinding the dried particles with a mortar and pestle. Finally, aqueous PNIPAM suspensions were prepared by adding PNIPAM powder in Milli-Q water. The conditions for the synthesis of the particles used in this work are tabulated in table 1.

	
	\subsection{2.2 Dyanamic Light Scattering}
	DLS was used to quantify the average sizes, size distributions, and thermoresponsivity of PNIPAM particles in dilute aqueous suspensions. The DLS experiments were performed using a Brookhaven Instruments Corporation (BIC) BI-200SM spectrometer attached with a 150 mW solid-state laser (NdYVO4, Coherent Inc., Spectra Physics) having an emission wavelength of 532 nm. The details of the setup is given elsewhere \cite{D. Saha_2014, D.Saha_2015}. A glass cuvette filled with the sample was held in a refractive index matching bath filled with decaline. The temperature of the sample was controlled by water circulation with a temperature controller (Polyscience Digital). A Brookhaven Instruments BI-9000AT digital autocorrelator was used to measure the intensity autocorrelation function $g^{(2)}(q,t)$ of the scattered light which is defined as $g^{(2)}(q,t) = \frac{<I(q,0)I(q,t)>}{<I(q,0)>^{2}} =  1+ A|g^{(1)}(q,t)|^{2}$ \cite{B. J. Berne_1975}. Here $q$, $I(q,t)$, $g^{(1)}(q,t)$ and $A$ are the scattering wave vector, the intensity at a delay time $t$, the normalized electric field autocorrelation function and the coherence factor, respectively. The scattering wave vector $q$ is related to the scattering angle $\theta$, $q=(4\pi n/\lambda)\sin(\theta/2)$, where $n$ and $\lambda$ are the refractive index of the medium and the wavelength of the laser, respectively. The decays of the normalized intensity autocorrelation functions,  $C(t) = \frac{g^{(2)}(q,t)-1}{A}$, measured for all the samples used in this work, were fitted to stretched exponential functions of the form $C(t) = [ \exp\{-(t/\tau)^\beta \}^2]$, where $\tau$  and $\beta$ are the relaxation time of the particle and the stretching exponent, respectively. The stretching exponent was used to calculate the mean relaxation time and the second moment of the relaxation time spectrum of the  particles in suspension using the relation, $<\tau> = (\frac{\tau}{\beta})\Gamma(\frac{1}{\beta})$ and $<\tau^{2}> = (\frac{\tau^{2}}{\beta})\Gamma(\frac{2}{\beta})$, respectively \cite{C. P. Lindsey_1980}, where $\Gamma$ is the Euler Gamma function. The mean hydrodynamic diameter $<d_{H}>$ of the particle can be calculated by using the Stokes-Einstein relation $<d_{H}>=\frac{k_{B}T<\tau>q^2}{3\pi\eta}$ \cite{B. J. Berne_1975, A. Einstine_1905}, where $k_{B}$, $T$ and $\eta$ are the Boltzmann constant, the absolute temperature and the viscosity of the solvent, respectively. The size distributions of PNIPAM particles in aqueous suspensions, obtained from DLS data using the CONTIN algorithm \cite{S. W. Provencher_1982, A. Scotti_2015}, were quantified by a dimensionless polydispersity index (PDI) which is defined as the ratio of the standard deviation or width of the particle size distribution, $\sigma = \frac{k_{B}T\sqrt{<\tau^{2}> - <\tau>^{2}}q^{2}}{3\pi\eta}$, and the average hydrodynamic diameter $(<d_{H}>)$ of the particle.   
	
	In the DLS experiments for particle-size measurements, very dilute suspensions ($0.01$ wt\%) of PNIPAM particles were prepared by adding dried PNIPAM powder in  5ml of Milli-Q water. The prepared suspension was filled in a glass cuvette and was placed in the sample holder of the DLS setup attached with the temperature controller. The thermoresponsive behaviour of aqueous suspensions of PNIPAM particles was characterized in the temperature range 18 - $50$$^{\circ}$C at intervals of $2$$^{\circ}$C after allowing an equilibration time of 15 min for each temperature during cooling and heating. The maximum swelling ratio $\alpha$ of PNIPAM particles in aqueous suspensions is defined as $d_{20^{\circ}C}/d_{50^{\circ}C}$ ($d_{fullswell}/d_{fullshrunk}$), where $d_{20^{\circ}C}$ and $d_{50^{\circ}C}$ refer to the hydrodynamic diameter of the particles at $20$$^{\circ}$ and $50$$^{\circ}$C ($d_{fullswell}$ and $d_{fullshrunk}$) respectively.     
	
	\subsection{2.3 Differential Scanning Calorimetry}
	
\begin{figure}[!t]
			\renewcommand{\figurename}{Figure}
			\includegraphics[width=5.5in]{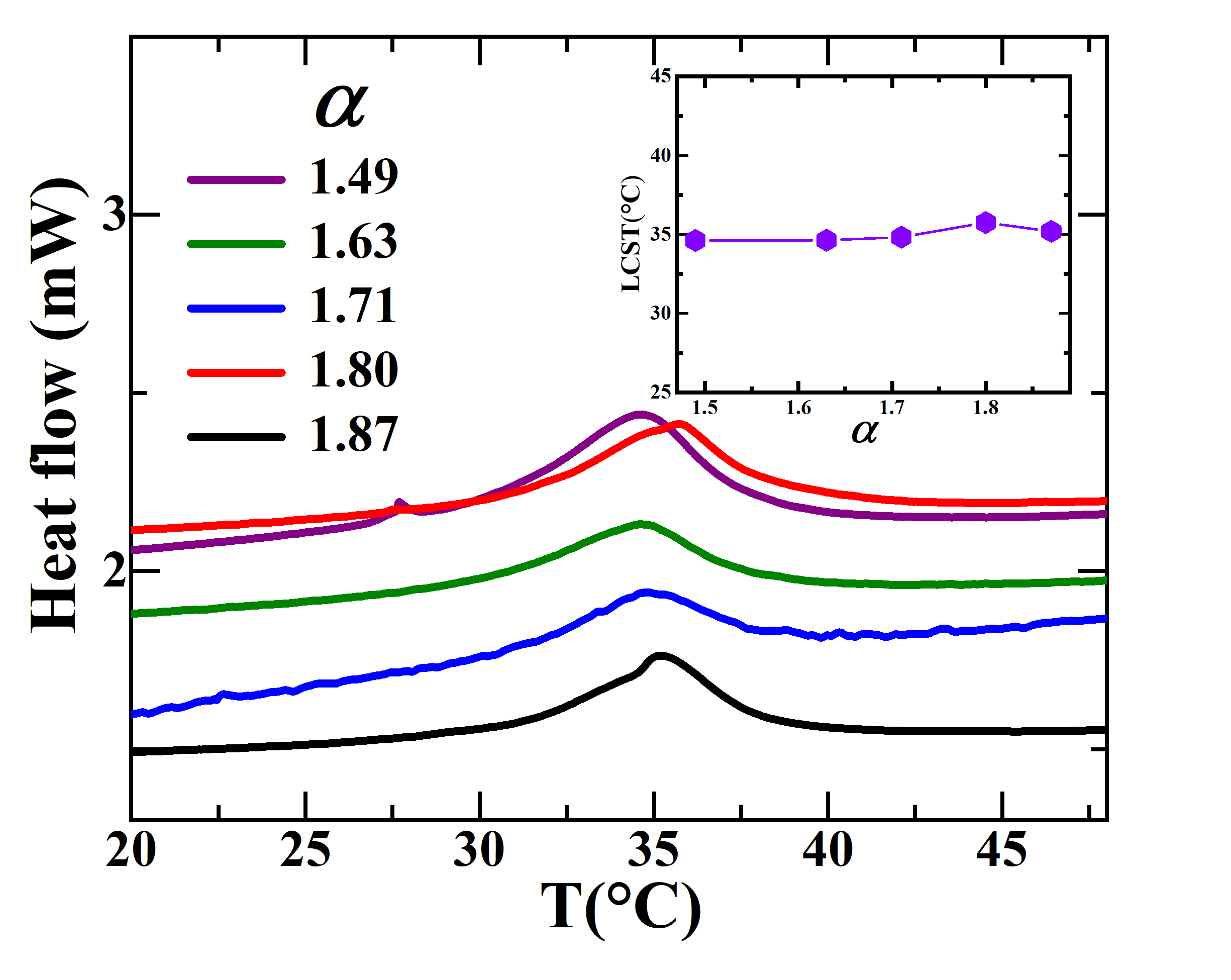}
			\caption{ Heat flow as a function of temperature (T) for aqueous PNIPAM suspensions of different maximum swelling ratios ($\alpha$) DSC. The inset shows the plot of the LCST {\it vs.} the maximum swelling ratio $\alpha$.}	
			\label{PNIPAM-DSC}
		\end{figure}

	The LCSTs of PNIPAM particles of different maximum swelling ratios in dense aqueous suspensions were estimated using DSC (Mettler Toledo, DSC 3). The heat flows estimated in the DSC experiments for aqueous PNIPAM suspensions of different maximum swelling ratios are plotted in figure \ref{PNIPAM-DSC} as a function of temperature T. The temperature corresponding to the endothermic peak in figure \ref{PNIPAM-DSC} is characterized as the LCST of the PNIPAM particles in the aqueous suspensions. The values of LCSTs for PNIPAM particles of different maximum swelling ratios are shown in the inset of figure \ref{PNIPAM-DSC}. It is clear from the inset of figure \ref{PNIPAM-DSC} that the LCSTs of PNIPAM particles in aqueous suspensions  are approximately the same for all the maximum swelling ratios observed in this research study.
	
	\subsection{2.4 Scanning Electron Microscopy}
	PNIPAM particle assemblies were visualized using a SEM, GEMINI column, ZEISS, Germany, with an accelerating voltage of 5 kV. Aqueous PNIPAM suspensions of concentration $0.01$ wt\% were poured on an indium tin oxide (ITO)-coated substrate using a pipette and were dried overnight at $25$$^{\circ}$C. The ITO substrate was then loaded on the SEM stage for imaging. The electron beam interacts with the atoms of the sample. The backscattered secondary electrons were used to produce the surface images of the samples. An image analysis software ImageJ was used to analyse the SEM images. Some clusters of two or three particles were found in the SEM images. These were avoided in the size analysis. The average particle sizes were estimated from around 1000 particles. 
	
	\subsection{2.5 Rheology}
	Rheological measurements were performed using a stress controlled Anton Paar MCR $501$ rheometer. Concentrated aqueous suspensions of PNIPAM particles were prepared by adding dried PNIPAM powder in Milli-Q water. The suspension was then stirred for 24 h and sonicated for 30 min. The concentrated suspension was next diluted to prepare different concentrations of PNIPAM suspensions. For the dilute samples, a double gap geometry (DG-26.7) having a gap of 1.886 mm and an effective length of 40 mm was used. For the concentrated suspensions, a cone-plate geometry (CP-25) having cone radius $r_{c} = 12.491$ mm, cone angle $\alpha = 0.979^{\circ}$, and measuring gap {\it d} = 0.048 mm was used. For the rheological measurements, a sample volume of $3.8$ ml was loaded in the DG-26.7 geometry, while a sample volume of $0.07$ ml was loaded in the CP-$25$ geometry. The temperature of the sample was controlled using a Peltier unit, and a water circulation system (Viscotherm VT2) was used for counter cooling. Silicon oil of viscosity 5cSt was used as a solvent trap oil to avoid solvent evaporation. The viscosity $\eta$ measurements of aqueous PNIPAM suspensions were performed by varying shear rates $\dot{\gamma}$ from $0.001$ to $4000$ $s^{-1}$. The  zero-shear viscosities $\eta_{0}$ of aqueous PNIPAM suspensions were estimated by fitting the $\eta$ {\it vs.} $\dot{\gamma}$ curves with the Cross model \cite{M. Cross_1965}:
	\begin{equation}
	\frac{\eta-\eta_{\infty}}{\eta_{0}-\eta_{\infty}}=\frac{1}{1+(k\dot{\gamma})^{m}}
	\end{equation}
	where $\eta_{0}$ and $\eta_{\infty}$ are the low and high shear rate viscosity plateaus respectively, $k$ is a time constant related to the relaxation time of the macromolecules in aqueous suspensions and $m$ is a dimensionless exponent. The relative viscosity $\eta_{rel}$ of an aqueous PNIPAM suspension is defined as the ratio of its zero-shear viscosity $\eta_{0}$ and the viscosity of water $\eta_{s}$: $\eta_{rel} = \eta_{0}/\eta_{s}$.
	
	The PNIPAM particles are soft and start deforming when packed above the random close packing volume fraction of undeformed monodisperse spheres ($\phi_{rcp} = 0.64$). Since the volume fraction $\phi$ is not appropriate for soft deformable particles, we have defined a modified parameter called the effective volume fraction $\phi_{eff}$. The effective volume fractions $\phi_{eff}$ of the aqueous PNIPAM suspensions were obtained from the relation $\phi_{eff} = nV_d$ \cite{J. Mattson_2009}, where $n$ is the number of particles per unit volume and $V_d = \pi(<d_H>)^3/6)$ is the volume of the undeformed particle of average hydrodynamic diameter $<d_H>$ in a dilute suspension. More details can be found in a previous contribution by author’s group \cite{S. K. Behera_2017}. The relative viscosity, $\eta_{rel}$, data is plotted {\it vs.} the polymer concentration, $c$, in figure 2. The data is fitted to the Batchelor's equation \cite{G. K. Batchelor_1977, J. F. Brady_1995} given by equation (2):

	\begin{equation}
	\eta_{rel}=1+2.5(c/c_{p})+5.9(c/c_{p})^{2}
	\end{equation}
	The polymer concentration inside each particle, $c_{p}$, is extrapolated from these fits. The effective volume fractions is next calculated using the relation $\phi_{eff}=c/c_{p}$.

\begin{figure}[!t]
			\renewcommand{\figurename}{Figure}
			\includegraphics[width=5.5in]{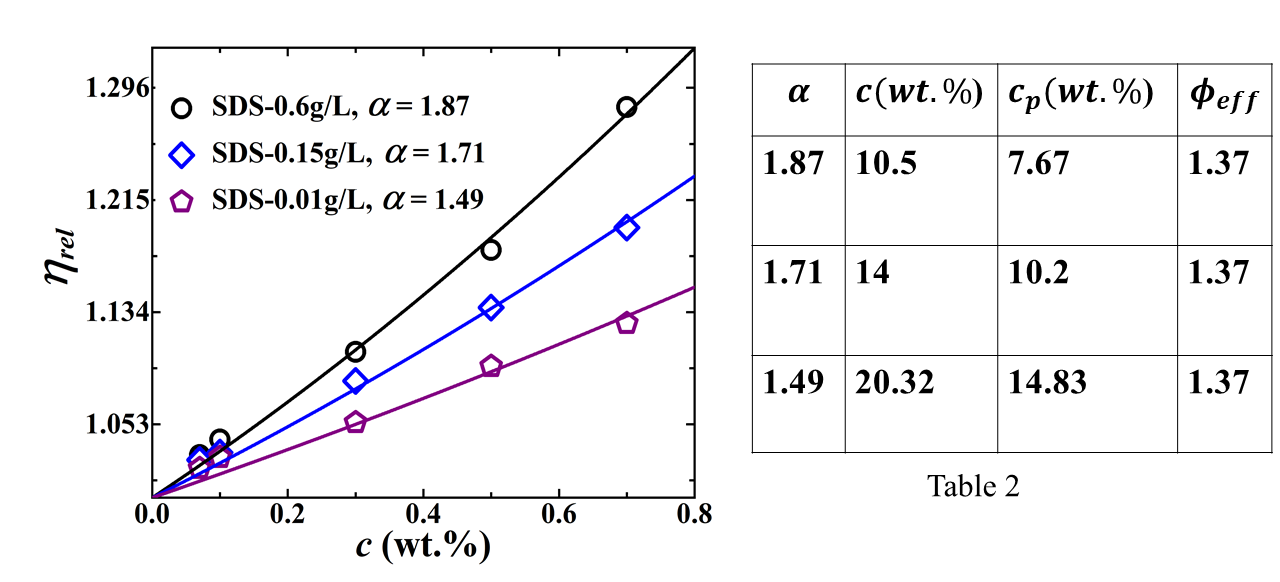}
			\caption{Relative viscosity $\eta_{rel}$ {\it vs.} concentration $c$ in wt\% for aqueous suspensions of PNIPAM particles of different swelling ratios $\alpha$. The solid lines are fits to equation (2). Table 2 shows the particle swelling ratios $\alpha$, the fitted value of the polymer concentration, $c_{p}$, inside each particle in the swollen state, and the effective volume fraction $\phi_{eff}$ of the particles in aqueous suspension.}	
			\label{PNIPAM-Rel_V}
		\end{figure}

	The relative viscosity $\eta_{rel}$ of dilute PNIPAM aqueous suspensions is plotted in figure \ref{PNIPAM-Rel_V} as a function of the particle concentration c for three different maximum swelling ratios $\alpha$ = 1.87, 1.71 and 1.49. The effective volume fraction of PNIPAM particles in an aqueous suspension at a particular concentration was estimated by extracting the parameter $c_{p}$ from the fits to the  Batchelor's equation (equation (2)), denoted by solid lines in figure \ref{PNIPAM-Rel_V}. The effective volume fraction of aqueous PNIPAM suspension at a particular concentration is related to the parameter $c_{p}$ by the relation $\phi_{eff}=c/c_{p}$. The estimated values of $c_{p}$,  the calculated effective volume fractions and the corresponding concentration of PNIPAM particles in aqueous suspensions for three different maximum swelling ratios 1.87, 1.71, and 1.49 are listed in table 2.
	
	\maketitle     
	%
	%

	\section{3. RESULTS AND DISCUSSIONS}
	
	The sizes ($<d_{H}>$), maximum swelling ratios $\alpha$, and PDIs of aqueous PNIPAM suspensions measured from DLS experiments at an angle $\theta=90^{\circ}$ are listed in table 1. The sizes of aqueous PNIPAM suspensions are controlled by varying the SDS concentration in a one-pot emulsion polymerization method. From the DLS data (table 1), it is observed that the sizes of PNIPAM particles in aqueous suspensions decrease with increase in SDS concentration. Particles of sizes 600-125 nm were synthesized by varying the SDS concentration from 0.01 g/L to 0.6 g/L. 
	
	Following the work of Hellweg {\it et al} \cite{T. Hellweg_2004}, we have characterized the swelling properties of the PNIPAM particles by defining a temperature-dependent swelling ratio $d_{T^{\circ}C}/d_{50^{\circ}C}$, where $d_{T^{\circ}C}$ and $d_{50^{\circ}C}$ are the particle diameters at temperatures T and ${50^{\circ}C}$, respectively. We measure the particle sizes as a function of temperature in temperature sweep DLS measurements. The thermoresponsive behaviour of PNIPAM particles in aqueous suspensions is shown in figure 3a. It is observed that the PNIPAM particles in aqueous suspensions swell maximally below $20^{\circ}$C and shrink fully above $45^{\circ}$C. It is  observed that the maximum swelling ratios defined as $d_{20^{\circ}C}/d_{50^{\circ}C}$ $d_{fullswell}$ and $d_{fullshrunk}$), were 1.49 and 1.87 for the PNIPAM particles of sizes 600 and 125 nm that were prepared using 0.01 g/L and 0.6 g/L concentration of SDS, respectively. Hence, the maximum swelling ratios of the PNIPAM particles in aqueous suspensions decrease with decrease in SDS concentration, which is accompanied by a simultaneous increase in particle size (table 1). The bigger PNIPAM particles in aqueous suspensions are therefore stiffer than the smaller particles. It is clear from the inset of figure 3a that the PNIPAM particles are thermoreversible in aqueous suspensions, since the particle diameters remain the same regardless of whether the measurement is performed during the heating or the cooling cycle.
	
	The size distributions of these particles measured from DLS data using the CONTIN algorithm are shown in figure 3b. The PDI values obtained from the DLS measurements for PNIPAM particles of different maximum swelling ratios are listed in table 1. A slight increase in the mean and width of the size distribution of PNIPAM particles in aqueous suspensions is observed with increase in particle size (figure 3b). Hence, from table 1 and figure 3b, it is concluded that particles of low PDIs can be synthesized by the one-pot emulsion polymerization method. The results can be explained by considering that higher concentrations of SDS in the one-pot method results in the presence of a larger number of charged species on the surfaces of the primary particles at different nucleation times. This results in the stabilization of the particles in solution without further agglomeration, thereby resulting in smaller particles. The further growth of chains or the coalescence of the primary nuclei at lower SDS concentrations leads to the observed wider distribution of particle sizes.

	The observed variation in maximum swelling ratios of the PNIPAM particles in aqueous suspensions (figure 3a) can be attributed to changes in the cross-linking density of the particles. A higher CLD can enhance the stiffness of the particles. The CLDs of the particles can be computed by estimating the Helmholtz free energies of the systems. The density of crosslinkers in the hydrogels is much lower in comparison to that in polymers. The effect of crosslinks on the suspension can therefore be neglected and the Helmholtz free energy can be written as the sum of the free energies due to the stretching of the network $W_{stretch}$ and due to the mixing of the polymer and the solvent $W_{mix}$ \cite{P. J. Flory_1942, P. J. Flory_1953, J. Li_2012}:

	\begin{equation}
	W=W_{stretch}(\lambda_{1},\lambda_{2},\lambda_{3}) + W_{mix}(\alpha)
	\end{equation} 
	where $\lambda_{1}$, $\lambda_{2}$, $\lambda_{3}$, the magnitudes of stretches in three-dimensional cartesian coordinates for a rectangular block of gel in the swollen state, arise from the swelling of the particle below the LCST. Above the LCST, PNIPAM particles are essentially composed of the dry polymer network.
	
	The stretching free energy is a function of the magnitudes of the stretches $\lambda_{1}$, $\lambda_{2}$, $\lambda_{3}$ and depends on the crosslink density (CLD), while the free energy due to the mixing of the polymer and the solvent is a function of the maximum swelling ratio $\alpha$. In an aqueous medium, the hydrogel imbibes the solvent and swells freely. Assuming that the hydrogel swells by the same stretch in all directions, the expressions for the Helmholtz free energy of mixing and the Helmholtz free energy due to stretching of the network can be written as \cite{S. Cai_2011}
	\begin{equation}
	W_{mix}=\frac{k_{B}T}{\Omega}\Big[(\alpha-1)log\Big(1-\frac{1}{\alpha}\Big)+\chi\Big(1-\frac{1}{\alpha}\Big)\Big]
	\end{equation}
	\begin{equation}
	W_{stretch}=\frac{1}{2}Nk_{B}T\Big[3\alpha^{2/3}-3-2log\alpha\Big]
	\end{equation}
	
	where $\Omega$ is the volume per water molecule, $N$ is the number density of the crosslinks which is the number of crosslinkers divided by the volume of the dry polymer and $\chi$, the dimensionless measure of the strength of the pairwise interactions between the species, is a function of $T$ and $\phi$.
	
	The Helmholtz free energy of the freely swelling hydrogel is finally written as the sum of the $W_{mix}$ and $W_{stretch}$ as a function of $\chi$, $\alpha$ and N.
	\begin{equation}
	W(\alpha,T)=\frac{1}{2}Nk_{B}T\Big[3\alpha^{2/3}-3-2log\alpha\Big]+\frac{k_{B}T}{\Omega}\Big[(\alpha-1)log\Big(1-\frac{1}{\alpha}\Big)+\chi\Big(1-\frac{1}{\alpha}\Big)\Big]
	\end{equation}
	At a particular temperature, the free energy will have a minimum, corresponding to the stable state (swollen or shrunken state) at equilibrium. Hence, the first order derivative of the free energy $W$ with respect to the maximum swelling ratio $\alpha$ should vanish: 
	\begin{equation}
	\frac{\partial W(\alpha,T)}{\partial \alpha}=Nk_{B}T\Big[(\alpha^{-1/3}-\alpha^{-1})+\frac{1}{N\Omega}\Big(log(1-\alpha^{-1})+\alpha^{-1}+\chi \alpha^{-2}\Big)\Big]=0
	\end{equation}  
	Equation (7) results in a useful relation between the CLD $N$ and the the maximum swelling ratio $\alpha$.
	\begin{equation}
	N\Omega=\frac{log(1-\alpha^{-1})+\alpha^{-1}+\chi\alpha^{-2 }}{\alpha^{-1}-\alpha^{-1/3}}
	\end{equation}
	
	The swelling of the PNIPAM latex is a measure of the CLD, and increases with reaction time \cite{X. Wu_1994}. This gives rise to a linear relation between the crosslinking density and the average hydrodynamic diameter of the PNIPAM particles: $N=k<d_{H}^{T}>$. Using this relation in equation (8), we get
	
	\begin{equation}
	<d_{H}^{T}>=d_{\alpha_{0}}\Big[\frac{log(1-\alpha^{-1})+\alpha^{-1}+\chi\alpha^{-2}}{\alpha^{-1}-\alpha^{-1/3}}\Big]
	\label{swell_ratio-diameter_particle}
	\end{equation}
	
	where $k$ is the proportionality constant establishing a relation between the crosslinking density of a particle with its size, $<d_{H}^{T}>$ is the diameter of the particle at a temperature $T$ and $d_{\alpha_{0}}$ is the size of the particle at a temperature $T$ for which $\alpha$ is the minimum.
	
	The mean hydrodynamic diameter <$d_{H}$> at $20^{\circ}$C {\it vs.} the maximum swelling ratio $\alpha$ is plotted in Figure 4. The solid line is a fit of the data to equation (9). The values of the fitting parameters $d_{\alpha_{0}}$ and $\chi$, obtained from fitting the data in figure 4 with equation (9), are $784\pm294$ nm and $0.60\pm0.11$ respectively. The value of $d_{\alpha_{0}}$ can be used to calculate the CLD per particle. The CLDs for different sizes of the particle at $20^{\circ}$C are calculated using the relation: $N = d_{T}/d_{\alpha_{0}}\Omega$ and are shown in table 3. From this table, it is clear that bigger particles have higher CLDs and hence are stiffer. This is because the higher SDS concentration stabilizes the primary particles at the early stage of the nucleation, thereby resulting in smaller particles and a lower CLD inside the particle. However, if the SDS concentration is low, a longer time is required to stabilize the particles, due to which we get larger sizes and higher crosslinker concentrations inside the particles.      

	The PNIPAM particles in dry state were observed using SEM. The SEM images of aqueous PNIPAM suspensions for different maximum swelling ratios are shown in figure 5a-e. The average sizes <$d_{H}$> of the PNIPAM particles, at $25^{\circ}$C, estimated from SEM images, are plotted in figure 5f {it vs.} the maximum swelling ratio $\alpha$. The error bars in  figure 5f correspond to the standard deviation of the particle sizes. It is seen that the particle size increases with decrease in the maximum swelling ratio $\alpha$ (figure 5f). A slight increase in the width of the distribution of the particle size is observed with decrease in $\alpha$. This observation agrees well with the DLS results reported earlier in this study. However, the sizes of the particles estimated from SEM images are found  to be less than those measured from DLS experiments at $25^{\circ}$C. This is because, below the LCST, the PNIPAM particles in suspension studied in the DLS measurements imbibe water from their envirenment. The particles visualized in the SEM images are, on the other hand, in the compeletely dry state. The DLS experiments therefore give larger values of <$d_{H}$> compared to the SEM images in which the particles are in a dried state.

	Temperature sweep experiments at a fixed oscillatory strain amplitude 0.3$\%$ and angular frequency 1 rad/s were performed to study the temperature dependence of the viscoelastic moduli $G'$ and $G''$. The temperature was changed at the rate of $1^{\circ}$C/min in both cooling and heating experiments. The temperature responses of the viscoelastic moduli $G'$ and $G''$ for aqueous PNIPAM suspensions of three different maximum swelling ratios 1.87, 1.71, and 1.49 at an effective volume fraction $\phi_{eff}$ = 1.37 are shown in figure 6a. It is clear from the inset of figure 6a that for the same effective volume fraction $\phi_{eff}$ = 1.37 below the LCST (T = $20^{\circ}$C), the elastic modulus $G'$ of the suspensions increases with decrease in maximum swelling ratio. It is observed that below the LCST, the elastic modulus $G'$ dominates over the viscous modulus $G''$, which indicates that the system is a  viscoelastic solid under these conditions. As the temperature increases, both $G'$ and $G''$ decrease. The difference between the two moduli $G'$ - $G''$,  which is a measure of the relative solidity of the viscoelastic suspension, is plotted in figure 6b. It is observed that the relative solidity gets smaller as the temperature increases and eventually becomes negligible near the LCST. This indicates that the system loses its rigidity near the LCST. For the soft particles ($\alpha$ = 1.87), the viscous modulus $G''$ dominates over the elastic modulus $G'$ near the LCST which indicates liquid-like behaviour of the system. This can be attributed to the sudden collapse in the sizes of PNIPAM particles near the LCST.  The temperature at which both $G'$ and $G''$ abruptly decrease is characterized as the LCST of the aqueous PNIPAM suspension. As the temperature is increased beyond the LCST, a gradual increase in the moduli is observed. The above result agrees well with the observations reported in an earlier work \cite{G. Romeo_2010}. It has been reported that the interaction between the PNIPAM particles in aqueous suspensions changes from attractive to repulsive at the LCST. The increase in both the moduli at temperature above the LCST is attributed to an increase in the attractive interactions between the PNIPAM particles in aqueous suspensions. This leads to the formation of gel networks which manifests as an increase in the moduli at these temperatures. The LCST of the PNIPAM particles of different maximum swelling ratios in aqueous suspensions, estimated from the rheological measurements, are plotted figure 6c. We note that the LCST values of PNIPAM particles in aqueous suspensions measured from temperature sweep rheology experiments match closely with the LCST values estimated from the DSC measurements.        
	
	figure 7a and b show the temperature dependence of  $G'$ and $G''$ while heating and cooling for PNIPAM particles in aqueous suspensions having maximum swelling ratios 1.87 and 1.49 respectively. It is observed that both the viscoelastic   moduli $G'$ and $G''$ follow the same paths during both cooling and heating cycles for suspensions of the stiffer PNIPAM particles of swelling ratio 1.49 (figure 7a). However, the viscoelastic  moduli for suspensions of the soft particles ($\alpha$ = 1.87) are observed to deviate from each other near the LCST while cooling (figure 7b). The observed behaviour in $G'$ and $G''$ for the soft PNIPAM particles in aqueous suspensions can be attributed to changes in the particle packing behaviour due to their higher thermoresponsivity and deformability.

	\section{4.Conclusions}
	Thermoresponsive PNIPAM particles of different sizes were synthesized by varying the SDS concentration in the emulsion polymerization method. Increasing the concentration of SDS produces smaller and relatively softer particles having lower PDIs. The stiffnesses of the particles are quantified in terms of their maximum swelling ratios. This study quantifies the variation of the maximum swelling ratios of the particles with their size and demonstrates that changes in the stiffness of the particles with changing size can be attributed to the different CLDs of the particles. The CLDs of PNIPAM particles in aqueous suspensions, estimated using a statistical model for the Helmholtz free energies of the hydrogel and written as a function of the particle maximum swelling ratio $\alpha$, is observed to increase with decrease in maximum swelling ratios. The viscoelastic moduli of these particles in aqueous suspensions at an effective volume fraction $\phi_{eff}$ = 1.37, measured from temperature sweep experiments, are observed to increase with decrease in swelling ratios at a temperature below the LCST (at $20^{\circ}$C). A decrease in both the moduli is observed with increase in temperature.  The temperature at which both $G'$ and $G''$ show a minimum value is characterized as the LCST of the particles in aqueous suspensions. The LCST values of PNIPAM particles in aqueous suspensions, measured using DSC, closely match those obtained in rheological measurements.
	
	Particle packing studies are important to understand the dependence of the colloidal glass transition on particle parameters such as their sizes, size distributions, stiffnesses, and their packing configurations \cite{S. K. Behera_2017, B. S. Martin_2012, M. Rey_2017, J. Mattson_2009, D. Paloli}. Since the properties of PNIPAM particles in aqueous suspensions can be controlled by tuning the temperature of the medium and the stiffnesses of the constituent particles, PNIPAM suspensions can potentially be used as multifunctional materials \cite{Y. Shi_2015}. The property of PNIPAM particles in aqueous suspensions to form aggregates above the LCST can be exploited to study the onset of film formation \cite{S. Ugur_2007, S. Schmidt_2010}. The swelling and deswelling properties of aqueous PNIPAM suspensions can be used in sensor devices \cite{M. R. Islam_2014, X. Li_2015}. Research involving the controlled synthesis and aggregation mechanism of PNIPAM particles in aqueous suspensions is therefore very important.       
	\section{Acknowledgements}
	We thank K. M. Yatheendran for his help with SEM iamging and K. N. Vasudha for her help with DSC measurements. We acknowledge the funding provided by Raman Research Institute.
	
\begin{table}
			\renewcommand{\tablename}{Table}
			\renewcommand\thetable{1}
			\begin{center}
				\begin{tabular}{ |c|c|c|c|c|c|c| }
					\hline
					$NIPAM$ & $MBA$ & $KPS$ & $SDS$ & $d_{25^{o}C}$ ($nm$) & $polydispersity$ $index(\%)$ & $\alpha(d_{fullswell}/d_{fullshrunk})$ \\
					\hline
					14 & $1.4$ & $0.56$ & $0.01$ & $600$ & $10.29$ & ${\approx} 1.49$ \\
					\hline
					14 & $1.4$ & $0.56$ & $0.06$ & $450$ & $9.15$ & ${\approx} 1.63$ \\
					\hline
					14& $1.4$ & $0.56$ & $0.15$ & $270$ & $8.31$ & ${\approx} 1.71$ \\
					\hline
					14& $1.4$ & $0.56$ & $0.3$ & $155$ & $8.96$ & ${\approx} 1.80$ \\
					\hline
					14& $1.4$ & $0.56$ & $0.6$ & $125$ & $8.68$ & ${\approx} 1.87$ \\
					\hline
					
				\end{tabular}
				\caption{Summary of synthesis conditions used for the preparation of latexes of PNIPAM by the one-pot emulsion polymerization method and the properties of the synthesized particles. The concentrations of monomer (NIPAM), crosslinker (MBA), initiator (KPS) and surfactant (SDS) are expressed in g/L. All the latexes were prepared in $200$ mL volume.}
				\label{table:emulsion-polymerization}
			\end{center}
		\end{table}
		
		\begin{figure}[!t]
			\renewcommand{\figurename}{Figure}
			\includegraphics[width=6.2 in]{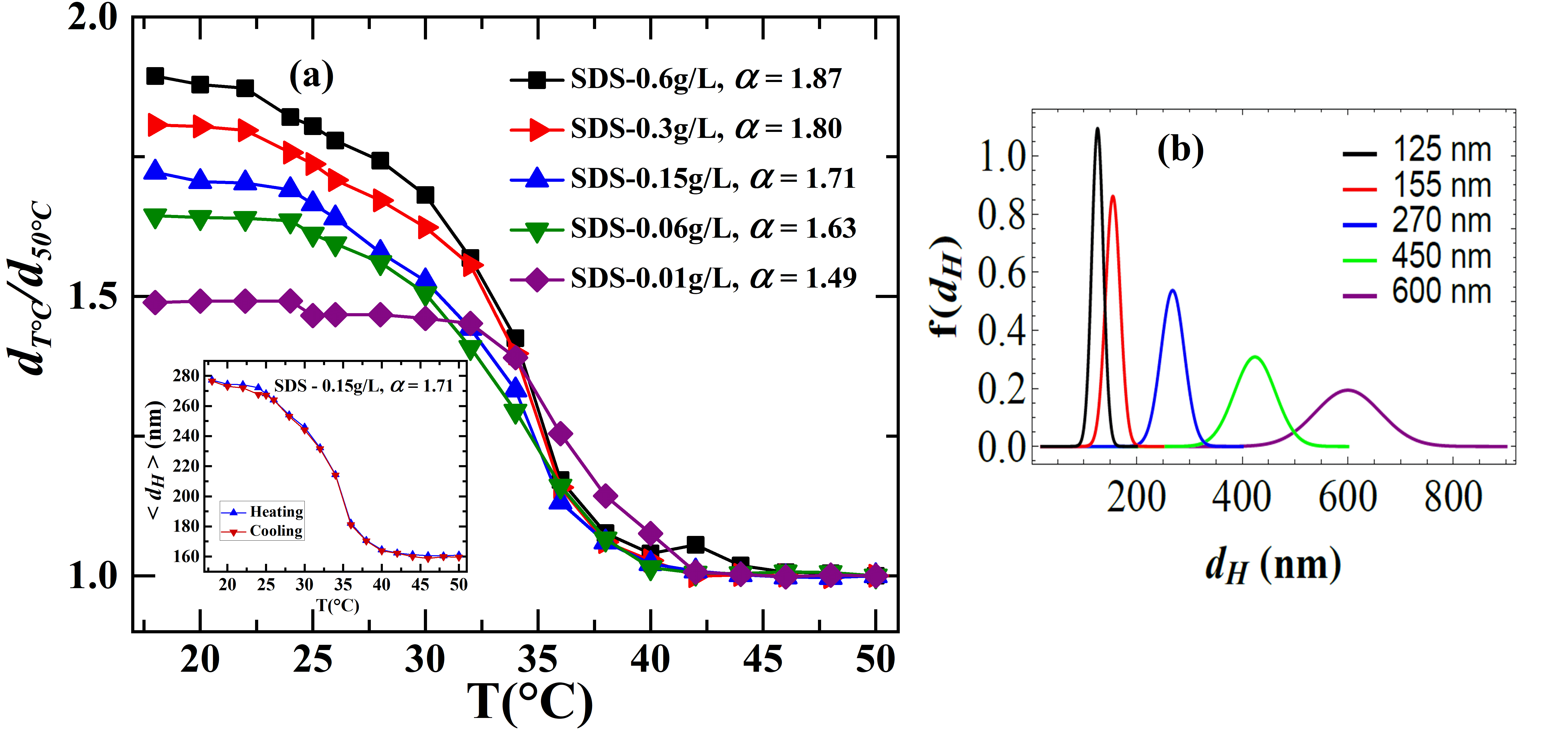}
			\caption{(a) Temperature-dependent swelling ratios $d_{T^{\circ}C}/d_{50^{\circ}C}$ as a function of temperature T of aqueous PNIPAM suspensions for different concentrations of SDS. In the inset, the average hydrodynamic diameter <$d_{H}$> of the PNIPAM particles of maximum swelling ratio 1.71 in aqueous suspension is plotted as a function of temperature while heating and cooling. (b) Particle size distributions of PNIPAM particles measured from DLS experiments for different particle sizes synthesized using various SDS concentrations in the one-pot emulsion polymerization method.}	
			\label{PNIPAM-DLS}
		\end{figure}
		\begin{figure}[!t]
			\renewcommand{\figurename}{Figure}
			\includegraphics[width=4.5in]{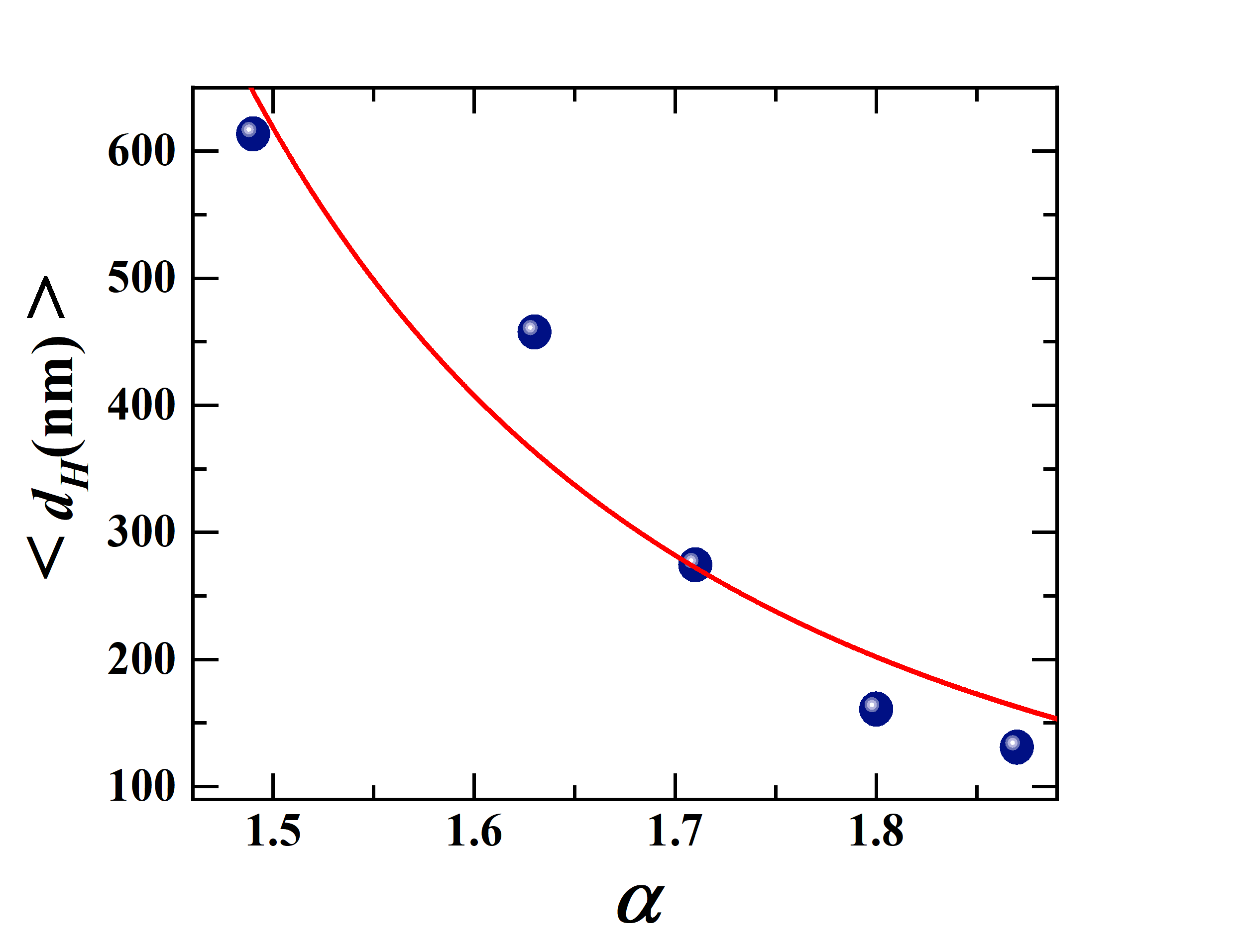}
			\caption{ Average hydrodynamic diameter <$d_{H}$> of the PNIPAM particles in aqueous suspensions at $20^{\circ}$C as a function of maximum swelling ratio $\alpha$. The solid line is a fit to equation (9).} 
			\label{PNIPAM-SR}
		\end{figure}
		\begin{table}
			\renewcommand{\tablename}{Table}
			\renewcommand\thetable{3}
			\begin{center}
				\begin{tabular}{ |c|c|c|c| }
					\hline
					$SDS$ & $d_{20^{\circ}C}$ & $\alpha$ & $N$ \\
					\hline
					$0.01$ $g/L$ & $613$ $nm$ & $1.49$  & $2.60\times10^{19}$ $m^{-3}$ \\
					\hline
					$0.06$ $g/L$ & $457$ $nm$ & $1.63$ & $1.94\times10^{19}$ $m^{-3}$ \\
					\hline
					$0.15$ $g/L$ & $274$ $nm$ & $1.71$  & $1.66\times10^{19}$ $m^{-3}$ \\
					\hline
					$0.3$ $g/L$ & $161$ $nm$ & $1.80$ & $6.83\times10^{18}$ $m^{-3}$ \\
					\hline
					$0.6$ $g/L$ & $131$ $nm$ & $1.87$ & $5.56\times10^{18}$ $m^{-3}$ \\
					\hline
				\end{tabular}
				\caption{Calculated CLD $N$ per particle from the fitting parameters $d_{\alpha_{0}}$ and $\chi$ for different sizes of the particles synthesized by varying the SDS concentration in the emulsion polymerization method. $d_{20^{\circ}C}$ and $\alpha$ are the hydrodynamic diameters  at $20^\circ$C and maximum swelling ratios $\alpha$ of PNIPAM particles in aqueous suspensions, respectively.. }  
				\label{table:Crosslink_density-particle_size}
			\end{center}
		\end{table}
		
		\begin{figure}[!t]
			\renewcommand{\figurename}{Figure}
			\includegraphics[width=6.2in]{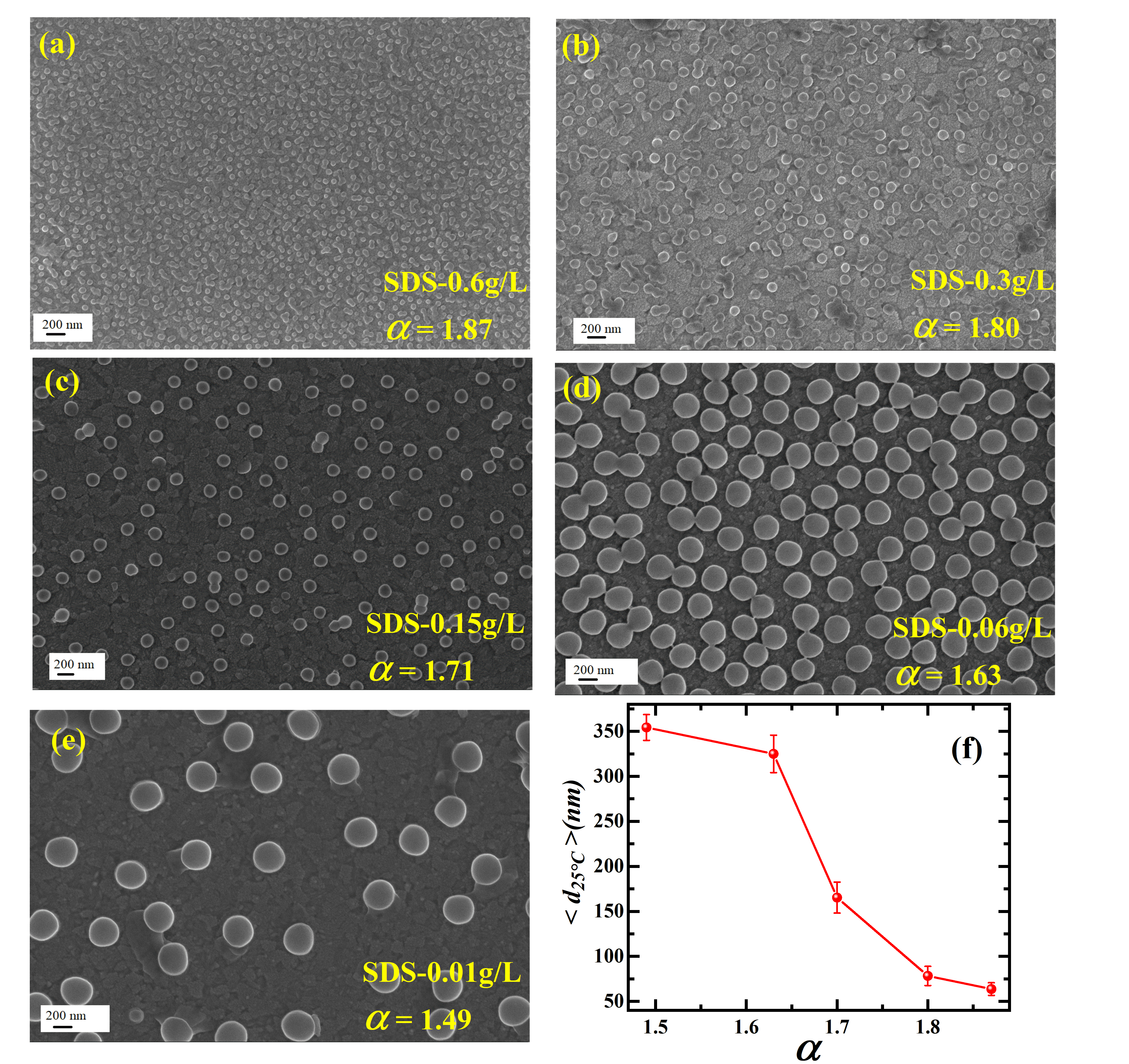}
			\caption{SEM images of PNIPAM particles at $25^{\circ}$C synthesized by emulsion polymerization method with different concentrations of SDS: (a) 0.6 g/L, (b) 0.3 g/L, (c) 0.15 g/L, (d) 0.06 g/L, and (e) 0.01 g/L. The scale bar corresponds to a length of 200 nm. (f) Shows the average diameter of the PNIPAM particles in aqueous suspensions estimated from SEM images as a function of maximum swelling ratios $\alpha$.}	
			\label{PNIPAM-SEM}
		\end{figure}
		\begin{figure}[!t]
			\renewcommand{\figurename}{Figure}
			\includegraphics[width=6.2in]{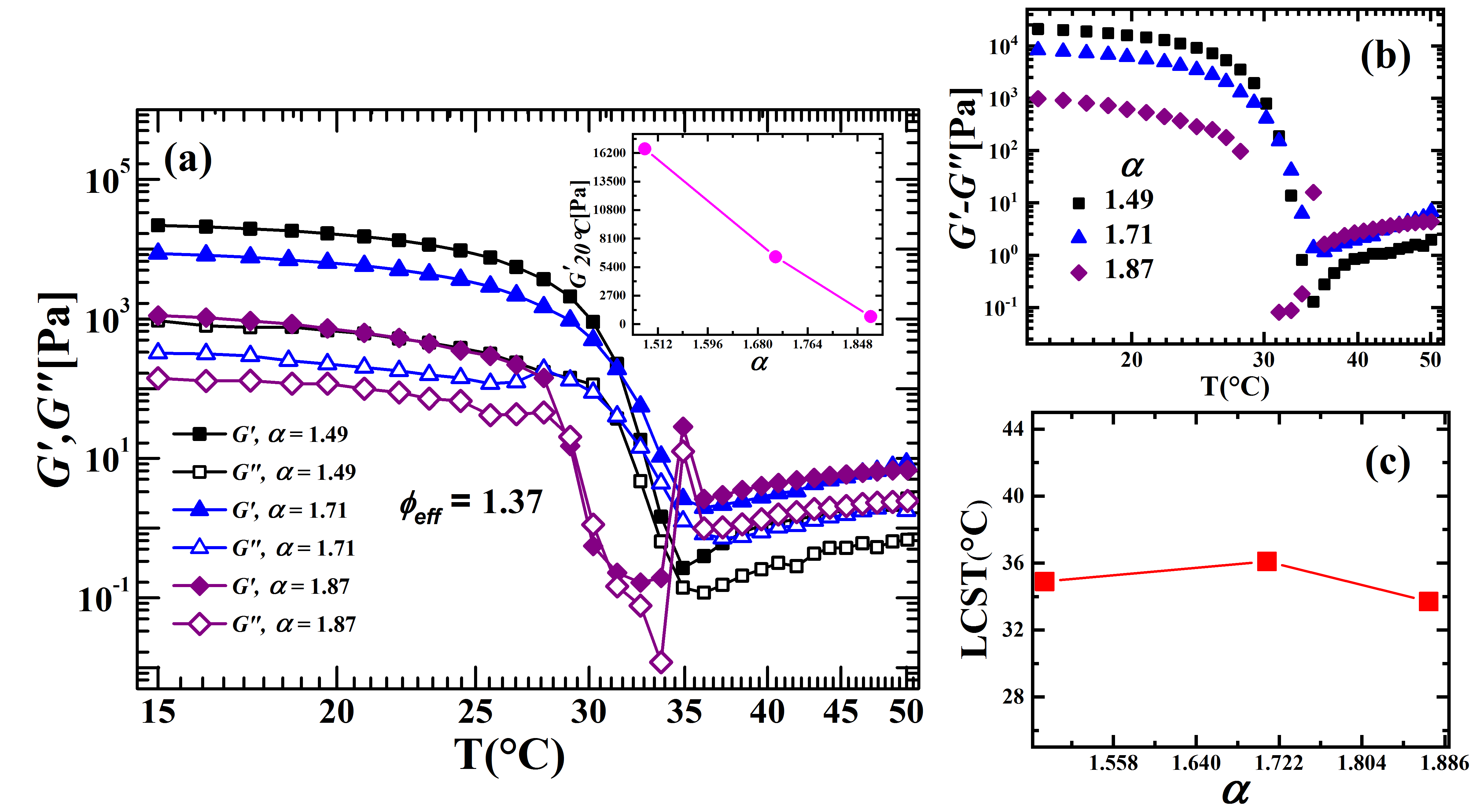}
			\caption{(a) Elastic modulus $G'$ (solid symbol) and viscous modulus $G''$ (open symbol) of aqueous PNIPAM suspensions of different maximum swelling ratios as a function of temperature T. (b) The difference between the elastic modulus and the viscous modulus, $G'$ - $G''$, {\it vs.} temperature T for aqueous suspensions of PNIPAM particles of different maximum swelling ratios $\alpha$. (c) Plot of the LCST, estimated from the rheological measurements, as a function of maximum swelling ratios $\alpha$.  }	
			\label{PNIPAM-TS}
		\end{figure}
		
		\begin{figure}[!t]
			\renewcommand{\figurename}{Figure}
			\includegraphics[width=5.5in]{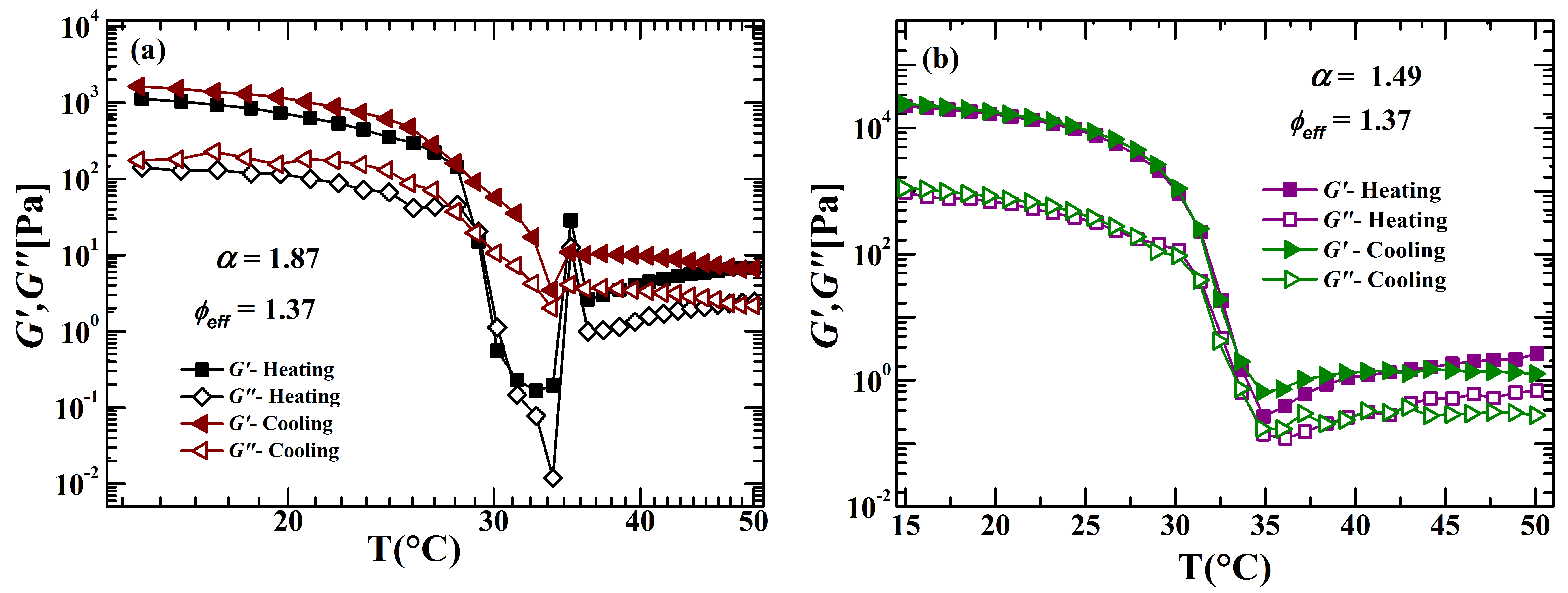}
			\caption{(a) The temperature-dependent storage modulus $G'$ and loss modulus $G''$ while heating and cooling a suspension of PNIPAM particles of swelling ratio = 1.87. (b) The temperature-dependent storage modulus $G'$ and loss modulus $G''$ while heating and cooling a suspension of PNIPAM particles of swelling ratio = 1.49. }	
			\label{PNIPAM-TS-2}
		\end{figure}

\end{document}